\documentclass[12pt,titlepage]{article}
\usepackage{epsfig}
\usepackage{graphicx}
\usepackage{amsmath}

\def\mp{{\widetilde m_P}}
\def\beq{\begin{equation}}
\def\eeq{\end{equation}}
\def\bea{\begin{eqnarray}}
\def\eea{\end{eqnarray}}
\def\bean{\begin{eqnarray*}}
\def\eean{\end{eqnarray*}}
\def\rh{{r_h}}
\def\Gfive{{G_{(5)}}}
\def\Gten{{G_{(10)}}}

\begin{document}

\title{Bose condensation and branes}


\author{{Brian P. Dolan}\\
{\em Department of Mathematics, Heriot-Watt University}\\ 
{\em Colin Maclaurin Building, Riccarton, Edinburgh, EH14 4AS, U.K.}\\
[5mm]
{\em Maxwell Institute for Mathematical Sciences, Edinburgh, U.K.}\\
[5mm]
{\em Email:} {\tt B.P.Dolan@hw.ac.uk}}

\maketitle

\begin{abstract}
When the cosmological constant is considered to be a thermodynamical
variable in black hole thermodynamics, analogous to a pressure, its
conjugate variable can be thought of as a thermodynamic volume for
the black hole.  In the AdS/CFT correspondence this interpretation
cannot be applied to the CFT on the boundary but, 
from the point of view of the boundary $SU(N)$ gauge
theory, varying the cosmological constant in the bulk is equivalent
to varying the number of colors in the gauge theory.  This interpretation
is examined in the case of $AdS_5\times S^5$, for ${\cal N}=4$ SUSY
Yang-Mills at large $N$, and the variable thermodynamically conjugate to 
$N$, a chemical potential for color, is
determined.  It is shown that the chemical potential in 
the high temperature phase of the Yang-Mills theory
is negative and decreases as temperature increases, as expected.
For spherical black holes in the bulk the chemical potential 
approaches zero as the temperature is lowered below the Hawking-Page
temperature and changes sign
at a temperature that is within one part in a thousand of the 
temperature at which the heat capacity diverges.

\bigskip
\noindent

\begin{flushright}
PACS nos: 04.60.-m; 04.70.Dy; 11.25.-w
\end{flushright}

\end{abstract}


\section{Introduction}

It was suggested in \cite{KRT} that in the presence of a cosmological
constant a black hole with mass $M$ should be viewed as a thermodynamic system
for which the mass is interpreted as the enthalpy, 
$M=H(S,P)$, with entropy $S$ and 
pressure $P=-\frac{\Lambda}{8\pi G_N}.$
The variable thermodynamically conjugate to $P$ would then have the
natural interpretation of a volume, \cite{KRT} \cite{BPD1},
\beq V=\left.\frac{\partial H}{\partial P}\right|_S, \eeq
although it does not in general have any a priori connection to a
notion of volume in the geometric sense \cite{CGKP}. 

This picture has been extensively investigated recently, for a review see
\cite{AKMS}, with most
work focusing on the case of negative $\Lambda$ (an exception being \cite{deSitter}).  A natural question to ask in this context is what the role of the
cosmological constant might be on the boundary CFT, at
finite temperature, in the AdS/CFT correspondence.
As $\Lambda$ is then related to the 
number, $N$, of branes in the bulk, and this translates to the number of colors 
in the boundary gauge theory, 
the variable thermodynamically conjugate to $\Lambda$ 
should behave as a chemical potential for color, \footnote{This should not 
be confused with a chemical potential
for gluons or quarks, the former should vanish just as for photons and the latter vanishes in the absence of baryon violating processes. The
chemical potential for color does not break supersymmetry.}
\cite{KRT} \cite{Compressibility} \cite{HeatEngines}.

The chemical potential $\mu$ is calculated 
in $AdS_5\times S^5$, with a black hole in $AdS_5$, corresponding to a finite temperature ${\cal N}=4$ SUSY gauge
theory on the boundary at large $N$, 
\cite{Maldacena}.  The Hawking-Page phase transition in the bulk is equivalent to a phase transition in the boundary gauge theory with a mass gap in the high
temperature phase, \cite{Witten}.
We show that, in the high temperature phase of the boundary theory, 
$\mu$ is negative and is a decreasing function of temperature, consistent with general expectations for a chemical potential,
\cite{Callen}.
Conversely if the Hawking temperature is lowered below that of
the Hawking-Page phase transition the chemical potential associated 
with a 5-dimensional asymptotically AdS spherical black hole 
can become zero or even positive.  Though the black hole solution is not the
one relevant for the physics in that phase, the temperature at which 
$\mu$ vanishes is very close, to within one part in a thousand, 
of the temperature at which the heat capacity of the black hole diverges. 

In the AdS/CFT 
correspondence the cosmological constant
is normally considered to be a fixed parameter that does not vary. In $AdS_5 \times S^5$ however 
it is not given a priori, it is just a parameter in a solution 
of 10-dimensional supergravity, and is no more fundamental to the theory 
than a black hole mass or any other parameter in the metric.
Indeed in some scenarios it can be dynamically determined by a scalar 
potential, \cite{FGPW} \cite{MS}.  
In \cite{GKP} the thermodynamic energy of the
boundary conformal field theory was calculated as a function of volume, temperature and $N$, but the bulk metric only has one parameter if $\Lambda$ is fixed,
and even allowing $\Lambda$ to vary gives only two parameters which is 
not enough to provide 
thermodynamic potentials as a function of three independent parameters,
that requires an extra assumption.
Our philosophy here is to trade the two geometric parameters in the $AdS_5$ black hole, $\rh$ and $L$ below, for two thermodynamic variables, which are taken to be entropy and $N^2$ in the micro-canonical ensemble.

\section{The chemical potential}

The line element for an asymptotically AdS Schwarzschild black hole in 5-dimensions is 
\beq
d s^2= -f(r) d t^2 + \frac{1}{f(r)} d r^2 + r^2 d^2\Omega_{(3)},
\eeq
where $d^2\Omega_{(3)} $ is the (dimensionless) line element on a compact 3-dimensional space $\Sigma_3$ and
\beq
f(r)=k - \frac{8  \Gfive M }{3 \pi r^2 } + \frac{r^2}{L^2}.
\eeq 
Here $L$ is the anti-de Sitter length scale, with $\Lambda=-\frac{4}{L^2}$,
and we shall consider the two cases  $k=1$ ($\Sigma_3$ a unit 3-sphere)
and $k=0$ ($\Sigma_3$ a 3-torus). 

The event horizon radius, $\rh$, is the largest root of $f(r)=0$, 
allowing $M$ to be expressed as a function of $\rh$ and $L$,
\beq
M=\frac{3 \pi \rh^2(k L^2 + \rh^2)}{8 \Gfive L^2}. 
\eeq

One must bear in mind however that, in the AdS/CFT correspondence, $\Gfive$ is itself a function of $L$ since
\beq \frac{1}{16 \pi \Gfive} = 
\frac{V_{S^5}}{16 \pi \Gten} = \frac{ \pi^2 L^5}{16 \Gten},\eeq
where $V_{S^5}=\pi^3 L^5$ is the volume of the 5-dimensional sphere with radius
$L$ and $\Gten$ is the (fixed) 10-dimensional Newton constant.
Hence
\beq
M=\frac{3 \pi^4 \rh^2 L^3 (k L^2 + \rh^2)}{8 \Gten}. 
\label{eq:MrhL}
\eeq
 
We now wish to express $M$ as a function of the entropy, $S$, and the
number of colors of the gauge theory, $N$, with $N$ large.
The Bekenstein-Hawking entropy of the black hole is
\beq
S=\frac{1}{4}\frac{A}{\hbar \Gfive}=
\frac{\pi^5 L^5  \rh^3}{2 \ell_P^8},
\label{eq:SrhL}
\eeq
where $A=2\pi^2 \rh^3$ is the \lq\lq area'' of the black 
hole (for $k=1$ the volume of a unit 3-sphere is $2\pi^2$
and for simplicity we have used the same value for the 3-torus, this is not 
necessary for $k=0$ but any other choice does not materially
affect the argument). 
$\hbar \Gfive = \frac{\hbar \Gten}{\pi^3 L^5}$, 
is the cube of the 5-dimensional Planck length and $\hbar \Gten = \ell_P^8$
where $\ell_P$ is the 10-dimensional Planck length, which is kept fixed throughout.

The other relation we need is \cite{Maldacena}
\beq
L^4= 4\pi g_s (\hbar\alpha^\prime)^2 N = \frac{\sqrt{2}N}{\pi^2} \ell_P^4.  
\label{eq:NrhL}
\eeq

Using (\ref{eq:SrhL}) and (\ref{eq:NrhL}) in (\ref{eq:MrhL}) gives the mass as
\beq
M(S,N)=\frac {3\,\mp}{4} 
\left\{k\left(\frac {S}{\pi} \right)^{\frac 2 3} N^{\frac{5}{12}}
+ \left(\frac {S}{\pi} \right)^{\frac 4 3} N^{-\frac{11}{12}}\right\},
\label{eq:MSN}
\eeq
with $\mp=\frac{\sqrt{\pi} \, m_P}{2^{1/8}}$ and
$m_P=\frac{\ell_P^7}{\Gten}$ the 10-dimensional Planck mass.

The Hawking temperature follows from the standard thermodynamic relation
\beq
T=\left.\frac{\partial M}{\partial S}\right|_N = 
 \frac {\mp}{2\pi}
\left\{k\left(\frac {S}{\pi} \right)^{-\frac 1 3} N^{\frac{5}{12}}
+2\left(\frac {S}{\pi} \right)^{\frac 1 3} N^{-\frac{11}{12}}
\right\},
\label{eq:T_Hawk}
\eeq
(with Boltzmann's constant set to one).  There is no magic here,
this is just a trivial re-writing of
Hawking's formula relating temperature to the surface gravity of the black hole,
\beq
T=\frac{ \hbar f'(\rh)}{4 \pi}= \frac{\hbar (k L^2+ 2\rh^2 )}{2 \pi \rh L^2}.
\eeq

Note that, at fixed $N$, the $k=1$ temperature as a function of $S$ 
has a minimum of $\frac {\sqrt{2} \mp}{\pi N^{\frac 1 4}}$.  For any value of
$T$ above this there are two values of $S$ with the same temperature,
giving large black holes and small black holes.  This minimum in $T$
corresponds to a divergence in the heat capacity, $C_N=T\left(\frac{\partial S}{\partial T}\right)_N$, small black holes have negative heat capacity and
are always unstable.

The Gibbs free energy is
\beq G(T,N)=M - TS = \frac{\mp}{4}
\left\{k\left(\frac{S}{\pi}\right)^{\frac{2}{3}} N^{\frac{5}{12}} - 
\left(\frac{S}{\pi}\right)^{\frac{4}{3}} N^{-\frac{11}{12}}\right\}.\eeq
For spherical black holes, with $k=1$,
this is negative for $N^2 < \frac{S}{\pi}$, corresponding to $L<\rh$,
and black holes in this regime are more stable than 5-dimensional AdS with thermal radiation at the same temperature, this is the Hawking-Page phase transition \cite{HawkingPage} which is distinct from the instability due to negative 
heat capacity mentioned above.
When $G$ is positive spherical black holes are susceptible to 
decay to pure AdS plus radiation and this happens at the Hawking-Page 
transition temperature
\beq T_*=\frac{3}{2\pi}\frac{\mp}{N^{1/4}}.\eeq
$k=1$ black holes stable are against Hawking-Page decay only or $T>T_*$. 

The derivative of $M$ with respect to $N^2$ will be interpreted as a
chemical potential for the number of colors, \cite{KRT} \cite{Compressibility} \cite{HeatEngines},
\beq
\mu:=\left.\frac{\partial M}{\partial N^2}\right|_S=
\frac {\mp} {32} 
\left\{5k\left(\frac {S}{\pi} \right)^{\frac 2 3} N^{-\frac{19}{12}}
-11 \left(\frac {S}{\pi} \right)^{\frac 4 3} N^{-\frac{35}{12}}\right\} 
\label{eq:mu}
\eeq
($N^2$ is used in the definition of $\mu$, rather than $N$, because all 
fields in the boundary ${\cal N}=4$ SUSY Yang-Mills theory are in the adjoint
representation of $SU(N)$).
$\mu$ is a measure of the energy cost to the system
of increasing the number of colors.

\section{Discussion}

For flat $k=0$ black holes $\mu$ is always negative,
but for spherical $k=1$ black holes it becomes positive when
\beq
N^2 > \left(\frac{11}{5}\right)^{\frac 3 2}\frac{S}{\pi}.
\label{eq:mubound}
\eeq
For a bosonic system the vanishing of the chemical potential would be 
a signal of Bose-Einstein condensation, for a fermionic system it is 
a signal that the exclusion principle is coming into play.
In terms of temperature the bound (\ref{eq:mubound}) is saturated at
a temperature some 6\% below $T_*$,
\beq
T_0=\frac{21}{2\pi \sqrt{55}} 
\frac{\mp} {N^{\frac {1} {4} }}= 0.944\, T_*.
\eeq
The condition that the black hole be stable against Hawking-Page decay to
AdS plus thermal radiation is, in geometric variables, $\rh>L$ and
the inequality (\ref{eq:mubound}) translates to
\beq \rh^2 <\frac {5} {11} L^2,
\label{eq:rhLmubound}
\eeq
putting positive $\mu$ into the region of phase space where black holes
are unstable against the Hawking-Page transition, {\it i.e.} in the
low temperature phase of the gauge theory. Such black holes are not necessarily
irrelevant though, for temperatures just below $T_*$ one expects black holes
to make their presences felt through thermal fluctuations and they
would also be important if the quark-gluon plasma were supercooled.

As stated earlier the Hawking-Page instability at $\rh=L$
is distinct from the instability due to negative heat capacity,
the latter is in the regime $\rh^2 < \frac {L^2} {2}$.
The singularity in heat capacity at
\beq T_\infty=\frac{\sqrt{2}\, \mp}{\pi N^{\frac 1 4}},\eeq
is the lowest value the $k=1$ black hole temperature (\ref{eq:T_Hawk}) 
can achieve and it is only marginally below $T_0$,
\beq T_\infty=0.999\,T_0.\eeq
It should be borne in mind however that the singularity in heat capacity
at $T_\infty$ is not a cusp, as $T$ cannot go below $T_\infty$:
as a function of $S$, with $N$ fixed, $T$ has two branches 
and $C_N$ is negative on the low $S$ branch and positive on the high $S$ branch.
 
For a classical gas $\mu$ is negative and becomes more negative as $T$
is increased, quantum effects become important when $\mu$ approaches zero 
and switches sign.  Indeed these are general properties of a chemical
potential \cite{Callen} and we see that they are satisfied in the present case.
In the high temperature phase, where the entropy per degree of freedom
$\frac{S}{ N^2}$ is large, equations  (\ref{eq:T_Hawk}) and (\ref{eq:mu}) 
show that indeed
\beq
\mu \approx -\frac{11 N^{\frac 3 4 }}{2 \mp^3} \left( \frac{\pi T}{2}\right)^4
\label{eq:highTmu}
\eeq
is negative and a decreasing function of $T$. 
(\ref{eq:highTmu}) is a strict equality for all $T$ when $k=0$.

An important parameter in the discussion of chemical potentials is the fugacity
\beq \xi = e^{\frac {\mu}{T}}.\eeq
$\xi<1$ in the classical
regime and tends to zero as $T\rightarrow\infty$, with quantum effects 
corresponding to $\xi\approx 1$. 
The figure below shows a plot of $\xi$ as
a function of the dimensionless temperature
\beq  t= 2\pi  \frac{ N^{\frac 1 4} T }{\mp}
\eeq
for $k=1$.
On the lower branch $\xi$ is less than unity and is a 
decreasing function of $T$ above $T_\infty$.  At fixed $T$, $\xi\rightarrow 0$ as $N\rightarrow\infty$
on the lower branch and $\xi\rightarrow 1$ on the upper branch, but
finite $t$ is possible in the regime of validity of the solution with 
$\frac{T}{\mp}<<1$ and $N>>1$ at the same time.
Note that as the curve is traced from the high $T$, 
low $\xi$ branch, through $T_*$, $T_\infty$ is encountered before $\mu=0$,
even though $T_0>T_\infty$.

\begin{figure}[!ht]
\label{fig:fugacity}
\centerline{\includegraphics[width=8cm]{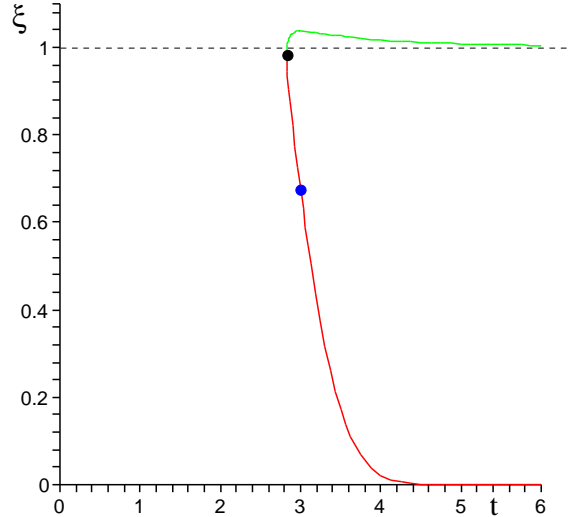}}
\caption{(Color online) fugacity as a function of $t$.
The blue (lower) dot denotes the Hawking-Page phase transition, where $t=3$
and $\xi=e^{-\frac{\pi}{8}}$: black holes are
stable on the branch below this point. The black (upper) dot is the point 
$T_\infty$ where
$C_P$ diverges, it lies marginally below $\xi=1$ (horizontal dotted line).
$C_P>0$ on the red (lower) branch of the curve and $C_P<0$
on the green (upper) branch.}  
\end{figure}

Bose-Einstein condensation and/or Fermion repulsion can viewed in terms
of flux density of the 5-form flux on $S^5$.
The density of flux on $S^5$ decreases as 
$n=\frac{N}{\pi ^3 L^5}\sim N^{-\frac 1 4}l_P^{-5}$ as $N$ is increased.
For large $N$ the flux is of course classical but as $N$ is decreased 
quantum effects will become important at some stage, when $\mu$ approaches zero.
If the flux is quantised and each unit of flux has a size $\lambda$
then the classical picture is only valid for $n<<1/\lambda^5$ or
$N>>\left(\frac{\lambda}{l_P}\right)^{20}$.

In the brane picture, when the event horizon is flat rather than spherical,
the first term in (\ref{eq:MSN}) is not present and the chemical potential
is always negative, which is a good thing as there is no Hawking-Page phase
transition in this topology, all such black branes, no matter how small $\rh$
is, are stable against decay to AdS plus thermal radiation.
The limit of $N$ co-incident branes can be obtained by first taking a stack
of branes with all adjacent branes having the same separation, $s$, and
then letting $s\rightarrow 0$. If the branes each have large mass $M_B$
then their Compton wavelength, $\lambda_B$, will be small.  We want to take
the limit $s\rightarrow 0$ and $M_B\rightarrow \infty$ and an important parameter characterising the physics is the ratio $\frac{s}{\lambda_B}$, with
$s\rightarrow 0$ and $\lambda_B\rightarrow 0$.  Provided this ratio is greater
than unity the brane wave functions do not overlap in the large $N$ limit
and quantum effects are not expected to be important, so we should
be taken with $s>>1$ for consistency.   
 
For simplicity, the discussion here has been restricted to maximally 
symmetric ${\cal N}=4$ Yang-Mills, in which the number of flavours is not
independent of the number of colors and all fields are in the adjoint of $SU(N)$.
It would be interesting to extend the analysis to ${\cal N}=1$
and varying $N_f$, by using classical solutions corresponding to less
symmetric situations such as finite temperature black hole versions of the 
${\cal N=1}$ solutions in \cite{MN} or \cite{CGR} for example.

If the gauge symmetry is broken, {\it e.g.} 
$U(N)\rightarrow U(N_1)\times U(N_2)$, then one could envisage extending 
the analysis to look for non-supersymmetric bulk gravity
solutions for which the black hole mass depends on $N_1$ and $N_2$ separately, generating two chemical potentials.
Thermodynamic equilibrium would then dictate the values of $N_1$ and $N_2$
in a similar manner to the way chemical equilibrium is achieved
in a system consisting of a molecular soup of different chemical species.

It is a pleasure to thank R.C.~Myers for useful conversations.

This research was supported in part by Perimeter Institute for Theoretical
Physics.  Research at Perimeter Institute is supported by the 
Government of Canada through Industry Canada and by the Province of Ontario
through the Ministry of Economic Development and Innovation.

\begin{appendix}

\end{appendix}

\end{document}